\documentclass{article}
\usepackage{spconf,amsmath,amssymb,graphicx}
\usepackage{multirow}
\usepackage{bbm}
\usepackage{booktabs}
\usepackage{float}
\usepackage{hyperref}
\usepackage{multicol, blindtext}
\usepackage{enumitem}
\usepackage{comment}

\def\l{{\mathbf l}}
\def\m{{\mathbf m}}
\def\x{{\mathbf x}}
\def\y{{\mathbf y}}
\def\z{{\mathbf z}}

\def\E{{\mathrm{\mathbf{E}}}}

\def\I{{\mathbf{I}}}
\def\N{{\mathbb{N}}}

\def\R{{\mathbb{R}}}
\def\Sw{{\mathbf{\Sigma_w}}}
\def\Sb{{\mathbf{\Sigma_b}}}
\def\X{{\mathbf X}}

\def\V{{\mathbf V}}
\def\Z{{\mathbf Z}}
\def\L{{\cal L}}
\def\Lprec{{ \mathbf{L} }}
\def\poster{{ \boldsymbol{\gamma} }}

\def\PHI{{\mathbf{\Phi}}}

\DeclareMathOperator{\Tr}{Tr}

\newcommand{\expnumber}[2]{{#1}\mathrm{e}{#2}}

\ninept

\title{Discriminative Training of VBx Diarization}

\name{
\begin{tabular}{c}
Dominik Klement$^1$, Mireia Diez$^1$, Federico Landini$^1$, Lukáš Burget$^1$, Anna Silnova$^1$, \\ \textit{Marc Delcroix$^2$, Naohiro Tawara$^2$}
\end{tabular}
}
\address{$^1$Brno University of Technology, Speech@FIT, $^2$NTT Corporation, Japan}

\begin{document}

\maketitle

\begin{abstract}
Bayesian HMM clustering of x-vector sequences (VBx) has become a widely adopted diarization baseline model in publications and challenges. It uses an HMM to model speaker turns, a generatively trained probabilistic linear discriminant analysis (PLDA) for speaker distribution modeling, and Bayesian inference to estimate the assignment of x-vectors to speakers. This paper presents a new framework for updating the VBx parameters using discriminative training, which directly optimizes a predefined loss. We also propose a new loss that better correlates with the diarization error rate compared to binary cross-entropy --- the default choice for diarization end-to-end systems. Proof-of-concept results across three datasets (AMI, CALLHOME, and DIHARD II) demonstrate the method's capability of automatically finding hyperparameters, achieving comparable performance to those found by extensive grid search, which typically requires additional hyperparameter behavior knowledge. Moreover, we show that discriminative fine-tuning of PLDA can further improve the model's performance. We release the source code with this publication.
\end{abstract}
\begin{keywords}
speaker diarization, VBx, clustering, variational Bayes, discriminative training
\end{keywords}

\section{Introduction}
\label{sec:intro}

Speaker diarization systems split audio recordings into speaker-homogenous segments and assign them speaker labels. Currently, three main approaches are used: clustering-based \cite{park2019auto,Landini2022VBx}, end-to-end \cite{horiguchi2020end,wang2023told}, or hybrid (combination of both)\cite{bredinpyannote2.1,delcroix23_interspeech}. While the latter two are the state-of-the-art \cite{wang2023told, delcroix23_interspeech}, clustering-based VBx diarization is competitive in wide-band multi-speaker domains \cite{delcroix23_interspeech}. Besides, VBx has been widely adopted in the community as a baseline both in publications \cite{wang2023told,raj2021multi} and in diarization challenges \cite{yu2022m2met, grauman2022ego4d, cheng2022conversational} as well as a reclustering stage in some systems \cite{horiguchi2021end,singh2023supervised}.

VBx is a clustering-based approach \cite{Landini2022VBx}, which requires as input a sequence of x-vectors \cite{snyder2018x} extracted from a uniformly segmented audio, and some corresponding initial labels, which are usually obtained by agglomerative hierarchical clustering (AHC). VBx further refines these speaker labels. The approach is based on a Bayesian hidden Markov model (HMM) where states correspond to speakers, and transitions correspond to speaker turns. 
Speaker-specific distributions are derived from a two-covariance probabilistic linear discriminant analysis (PLDA) model \cite{brummer2010spkpartitioning}, pre-trained on a large number of speaker-labeled x-vectors. 
The model uses variational Bayes (VB) inference \cite{Bishop2006} to obtain the speaker distributions and the refined speaker labels.

Being a clustering-based method, different approaches and objectives are used to train different parts of the model: the x-vector extractor is trained discriminatively with a speaker classification objective; the PLDA is trained generatively to maximize the likelihood of its training set, and VBx hyperparameters are selected by grid search to optimize diarization performance.
Alternatively, end-to-end systems present a single neural model trained discriminatively for diarization but they require large amounts of (speaker-labeled) conversational speech data, which are scarce. Thus, artificially simulated training sets are built. However, they hinder system performance and are a field of research on its own \cite{yamashita2022improving,landini2022InterDataEEND}. On the contrary, while the independent training of models used in clustering-based methods is seen as a downside, it allows for the usage of large amounts of real (single) speaker data to build robust speaker-discriminant models.

In the context of speaker recognition, discriminative training has been previously applied to generative models. In \cite{burget2011discriminatively}, a PLDA-like functional was proposed for evaluating the speaker verification scores, and its parameters were discriminatively trained using an objective function that directly addressed the speaker verification task, providing significant performance gains.
In \cite{McCree2014MulticlassDT}, an i-vector classifier was discriminatively trained using maximum mutual information to directly optimize the multi-class calibration criterion, to remove the need of using later a calibration backend.

In this paper, we propose a discriminative training framework for VBx (DVBx). 
With DVBx, all stages and parameters of the diarization pipeline can be jointly fine-tuned in an end-to-end fashion to minimize the diarization error directly.
Compared to standard EEND approaches, which are usually trained from scratch on simulated speaker conversations, VBx uses an x-vector extractor trained on a large number of real speaker utterances, and a PLDA model trained on a large number of such x-vectors. These components provide the model with a strong speaker discriminative power and serve as a well-founded starting point for DVBx fine-tuning.

While we envision DVBx as a method capable of state-of-the-art results, the main focus of this paper is to introduce the methodology and present fair comparisons with the well-established original VBx.
It is out of the scope of this paper to compare to current state-of-the-art approaches, which would require the use of VAD and overlap detection, and which are not used in the current work. 
In later sections, we will outline how we plan to extend the approach in directions that will produce results on par with end-to-end models.

To prove the validity of the method, we present a set of experiments that automatically find the set of VBx hyperparameters. 
Their tuning is usually done by an extensive, time-consuming grid search. Moreover, one cannot search through all possible combinations 
and has to restrict the parameter search space to a finite pre-defined set of values, which may lead to finding a suboptimal solution. 
We show that by automatically optimizing the hyperparameters only, we can find a setting achieving comparable performance to the baseline VBx model with parameters found by grid search \cite{Landini2022VBx} on three real datasets: AMI, CALLHOME, and DIHARD~II. 
Moreover, we show that the performance can improve even further by fine-tuning the PLDA parameters. All our code is publicly available \url{https://github.com/BUTSpeechFIT/DVBx}.

\section{Model Description}
\label{sec:model_description}
In this section, we briefly describe the necessary parts of the VBx model closely related to our work and refer the reader to the original VBx publication \cite{Landini2022VBx} for more details and the full description of the VB inference, the derivation of the update formulae or the definition of the hyperparameters.

VBx uses a Bayesian HMM to model the sequence of $d$-dimensional x-vectors $\hat{\X} = (\hat{\x}_1, \hat{\x}_2, ..., \hat{\x}_T)^\top \in \R^{T \times d}$ representing $T$ consecutive short speech segments in an input conversation.
To solve the diarization task, we need to infer the sequence of HMM latent variables $\Z = (z_1, z_2, ..., z_T)^\top \in \N^T$ representing the hard assignment of x-vectors to the HMM states, where each HMM state corresponds to one speaker in the conversation.
The state transition probabilities are modeled as
\begin{equation}
\label{eq:transition_probability}
    p(z_t = s | z_{t-1} = s') = (1-P_l)\pi_s + \delta(s = s')P_l,
\end{equation}
where $\delta$ refers to the Kronecker delta function, $P_l$ is the probability of staying in the same state, and $\pi_s$ is the probability of switching to state/speaker $s$. The HMM state distributions (i.e., distributions of the x-vectors corresponding to each speaker) are Gaussian with specific mean and shared within-speaker covariance matrix $\Sw$. VBx uses a Bayesian setting where the speaker means are latent variables
with Gaussian prior with global mean $\m$ and between-speaker covariance $\Sb$.
The parameters $\m$, $\Sb$, and $\Sw$ are taken from a pre-trained PLDA model.
For mathematical convenience, we first center the original x-vectors around the global mean $\m$ and transform them by LDA-like linear transformation $\E$ to obtain x-vectors $\X = (\hat{\X} - \mathbbm{1}\m)\E$ with zero global mean, diagonal between-speaker covariance matrix $\PHI$ and identity within-speaker covariance matrix. The matrices $\E$ and $\PHI$ can be obtained by solving the generalized eigenvalue problem $\Sb\E = \Sw\E\PHI$. For the transformed x-vectors, the HMM state distributions are 
$$p(\x_t | z_t=s) = \mathcal{N}(\x_t; \V\y_s, \I),$$
where $\V=\PHI^{\frac{1}{2}}$ and $\y_s$ is the state/speaker specific latent variable with standard normal prior.

For each input conversation, VBx uses VB inference~\cite{Bishop2006} to iteratively update the approximate posterior distributions $q(\y_s)$, responsibilities $\gamma_{ts} = q(\z_t = s)$ and the speaker priors $\pi_s$. After convergence, the responsibilities $\gamma_{ts}$ represent the (probabilistic) assignment of x-vectors to HMM states/speakers and therefore provide the solution to the diarization problem.

The approximate posteriors $q(\y_s) =\mathcal{N}(\y_s; \boldsymbol{\alpha}_s, \Lprec^{-1}_s)$ are updated as:
\begin{align}
\boldsymbol{\alpha}_s=\frac{F_A}{F_B}\mathbf{L}_s^{-1}\sum_t \gamma_{ts}\boldsymbol{\V^{\top}\x}_t,\;\;\;\;\;\;\mathbf{L}_s=\mathbf{I}+\frac{F_A}{F_B}\left(\sum_t \gamma_{ts}\right)\boldsymbol{\Phi},\nonumber
\end{align}
where $F_A$ and $F_B$ are model hyperparameters (described below). The responsibilities $\gamma_{ts}$ are updated using the standard (non-Bayesian) HMM forward-backward algorithm, except that the usual HMM state log-likelihood $\log p(\x_t | z_t)$ is replaced by a more complex term accounting for the uncertainty encoded in $q(\y_s)$:
\begin{equation}
    \begin{split}
         \log \overline{p}(\x_t|z_t) 
        & \propto F_A \left[ \boldsymbol{\alpha}_s^{\top}\V\x_t - \frac{1}{2} \Tr(\PHI[\mathbf{L}_s^{-1} + \boldsymbol{\alpha}_s\boldsymbol{\alpha}_s^{\top}])
        \right].\nonumber
    \end{split}
\end{equation}
Maximum likelihood type II  
updates are used for $\pi_s$ (see equation (24) in \cite{Landini2022VBx}).

Scalars $F_A, F_B$ are model hyperparameters that were introduced to 
control the VB inference. 
The acoustic scaling factor $F_A$ counteracts the independence assumption between observations. The speaker regularization coefficient $F_B$ can be used to control the number of output speakers (a high $F_B$ results in keeping fewer speakers).

The responsibilities $\gamma_{ts}$ for the first VB iteration are typically initialized from labels obtained by preceding AHC. Let $\hat{\boldsymbol{\gamma}}_t$ be a one-hot vector encoding the hard assignment of a speaker label to x-vector $\x_t$ as provided by AHC. Rather than using elements of such vectors directly as the initial responsibilities, we apply label smoothing:
\vspace*{-1mm}
\begin{equation}
    \label{eq:label_smoothing}
    \boldsymbol{\gamma}_t = softmax(\tau \cdot \hat{\boldsymbol{\gamma}}_t),  
\end{equation}
where a lower value of hyperparameter $\tau$ makes the initial assignment of x-vectors to speakers more uncertain.

\section{Discriminative Training}
Many diarization datasets have training and/or development data that can be used for tuning parameters. 
Conventional VBx uses an already trained (and fixed) x-vector extractor and PLDA model, and such data are only used to tune its hyperparameters.
In contrast, DVBx can utilize the same data to train all VBx parameters discriminatively. It is possible because VB update equations involve mathematical operations through which we can backpropagate loss gradients w.r.t model parameters, and the iterative VB inference can be unfolded, similarly to recurrent neural networks.
In this section, we describe the details of the DVBx setup.

\subsection{Losses}
Similar to many end-to-end diarization approaches \cite{horiguchi2020end,wang2023told}, the model is optimized using permutation invariant training (PIT) scheme \cite{yu2017pit}:
\begin{equation}
    \label{eq:pit}
    \L = \frac{1}{TS} \min_{\phi \in perm(S)} \sum_{t=1}^T H(\poster_t^{\phi}, \hat{\l}_t),
\end{equation}
where $perm(S)$ is the set of all permutations of sequence (1, 2, \dots, S), $\poster_t^{\phi}=(\gamma_{t 1}^{\phi}, \gamma_{t 2}^{\phi}, \dots, \gamma_{t S}^{\phi})^{\top} \in \langle 0, 1 \rangle ^S$ represents the permuted responsibilities predicted by the VBx model for $t$-th frame, and $\hat{\l}_t=(\hat{l}_{t1}, \hat{l}_{t2}, \dots, \hat{l}_{tS})^{\top} \in \{0, 1\}^S$ are the ground truth labels indicating speaker activities of all $S$ speaker in the $t$-th frame. For the loss function $H$, we first considered binary cross entropy (BCE) as normally used for end-to-end diarization system training:

\vspace*{-2mm}
\begin{equation}
    H_{B}(\poster_t^{\phi}, \hat{\l}_t) = \sum_{s=1}^S -\hat{l}_{ts}log(\gamma_{ts}^{\phi}) - (1-\hat{l}_{ts})log(1-\gamma_{ts}^{\phi}).
\end{equation}

However, VB inference is known to underestimate latent variable uncertainty, which leads to overly confident (close to either 0 or 1) responsibilities $\gamma_{ts}$. Consequently, the BCE loss becomes very high, as it heavily penalizes confident errors.
Overconfidence can adversely affect the performance of DVBx, as the loss primarily drives parameter updates to mitigate overconfidence rather than improving diarization performance. To compensate for such behavior, we optionally calibrate the responsibilities by passing them through the same function as Eq.~(\ref{eq:label_smoothing}) right before calculating the loss. The corresponding parameter $\tau_{calib}$ is also trained to ``please'' the BCE loss.

Despite BCE being the standard loss for training diarization systems \cite{fujita2019end}, it does not directly correlate with the diarization error rate (DER) (see Section \ref{sec:results}). Therefore, we propose expected detection error loss (EDE):
\begin{equation}
    H_{E}(\poster_t^{\phi}, \hat{\l}_t) = \sum_{s=1}^{S} (1-\gamma_{ts}^{\phi})\hat{l}_{ts} + \gamma_{ts}^{\phi}(1-\hat{l}_{ts}),
\end{equation}
which calculates expected miss, if $\hat{l}_{ts} = 1$, and expected false alarm, if $\hat{l}_{ts} = 0$. The main difference between EDE and the DER metric is that EDE operates with probabilities rather than hard decisions, and it double-counts speaker confusion frames, as confusion can be viewed as both a miss and a false alarm from the perspective of different speakers.
Note that EDE is not as sensitive to overconfident responsibilities as BCE; hence, it does not require calibration. However, if also calibrated by Eq. \eqref{eq:label_smoothing}, the larger the positive calibration constant $\tau_{calib}$,  the closer the EDE is to (hard) detection error due to over-saturated responsibilities.

\subsection{Backpropagation through VB Iterations}
The straightforward way to backpropagate through VBx is to run VB inference for a predefined number of iterations, obtain the refined speaker labels after the last iteration, compute the loss, and backpropagate the gradients. However, this leads to vanishing gradients when using a high number of VB iterations. 

As a solution, we compute the loss of Eq. (\ref{eq:pit}) using the responsibilities after each VB iteration and average them to obtain the final loss value. It is mathematically equal to averaging the gradients across the VB iterations, which speeds up the model convergence due to easier gradient propagation.

\subsection{Parameters}
The following VBx parameters were considered for fine-tuning in this publication. In some cases, re-parametrizations were used to ensure they remain valid.
$F_A$ and $F_B$ are trained directly. To ensure $P_l$ is a valid probability, we optimize $\mathrm{logit}(P_l)$ and apply sigmoid afterward. To ensure positive $\tau$, we train $\mathrm{\log}(\tau)$ (the same for $\tau_{calib}$).
Regarding PLDA parameters, the transformation matrix $\E$ is trained directly, and for the diagonal covariance matrix $\PHI$, we train the log of the diagonal.

\vspace{-2mm}
\section{Experimental Setup}
\label{sec:typestyle}

\vspace{-1mm}
\subsection{Data}
The proposed discriminative training does not require large amounts of training data, as only a few scalars are optimized, and the PLDA model is generatively pre-trained, so the discriminative training only fine-tunes its parameters.

To evaluate the method's performance, we use three datasets: DIHARDII~\cite{DIHARD19}, CALLHOME~\cite{Callhome}, and AMI Beamformed~\cite{carletta2005ami}. 
DIHARD II (DH) comes with a pre-defined split into two parts: development and testing. The development part is further split into two equal-sized speaker-disjoint (if possible) parts for training (8 hours) and validation (10 hours). For CALLHOME (CH), we use the well-adopted split\footnote{\href{https://github.com/BUTSpeechFIT/CALLHOME_sublists}{https://github.com/BUTSpeechFIT/CALLHOME\_sublists}} into two partitions: part1, part2, containing 249 and 250 files respectively. We further split part1 into two subsets: train (127 files - 4 hours) and validation (122 files - 3.5 hours), such that each part contains approximately the same number of files per each number of speakers (2-7). 
Lastly, we use the Full-corpus-ASR partition \cite{Landini2022VBx} (train set 65 hours, dev set 7.5 hours) in the case of AMI.

\subsubsection{Labels}
Like most standard clustering-based methods, VBx does not handle overlaps. Therefore, designing the ground truth (GT) labels of the overlap segments for discriminative training is not straightforward. First, we considered assigning each x-vector the GT label of the dominant speaker in the segment, as this is what the VBx system can estimate in the optimal case. Next, we considered the GT labels containing the values $p \in \langle 0,1\rangle$ representing the proportion of each speaker's speech to the total amount of speech in the segment.
In our preliminary experiments, the latter approach outperformed the former one in multiple training settings, so we decided to use it for the experiments presented in this paper.

\subsection{Model}
The model takes $b$ (batch size) pre-extracted x-vector sequences, extracted by ResNet101 \cite{he2016deep, zeinali2019but}, from $b$ utterances and processes them in parallel. Each concurrent instance runs VBx on a single utterance (i.e., we do not require equal-length utterances in a single batch) and computes the gradients w.r.t. the trainable parameters. Subsequently, gradients from parallel instances are averaged, and the trainable parameters are updated accordingly.

\subsubsection{Training Setup}
We experimented with two training strategies: a) training the hyperparameters ($F_A, F_B, \tau, P_l$) and then PLDA with hyperparameters fixed, and b) joint training of all parameters. 
If all are jointly trained, as PLDA has orders of magnitude more parameters than the number of hyperparameters, PLDA fine-tuning quickly decreases the loss but prevents hyperparameters from reaching optimal values. Therefore, we use the two-stage training strategy.

At the beginning of the inference, $F_A$ and $F_B$ are set to their theoretically correct value of 1. Furthermore, $\tau_{calib}$ default value is also set to 1. Since there are no theoretically correct values for $P_l$ and $\tau$, we considered several initial values, all of which had almost no effect on the model performance. Therefore, we set their values to the ones selected for the original VBx \cite{Landini2022VBx}.

VB inference is run for 10 iterations. We set $b=8$ and train the model for 500 epochs with Adam optimizer \cite{adam_optim}. 

In the VBx model, the trainable parameters have different dynamic ranges, and small changes in some of them have a strong impact on performance. This could be easily observed as gradients w.r.t. $F_A$ were much higher than w.r.t. other hyperparameters.
Therefore, our model required searching for suitable learning rates (LR) for the different parameters.
LR was set to $\expnumber{5}{-4}$ for $F_A$ and to $\expnumber{1}{-2}$ for the rest of the hyperparameters. For PLDA matrices with many more trainable parameters, the LR was set to $\expnumber{1}{-3}$.

Experiments showed that $P_l$ converged to 0 independently of its initial value. Moreover, we observed that training with $P_l=0$ throughout the whole training process did not significantly decrease the model performance. Therefore, we set $P_l=0$ in all DVBx experiments. This further simplifies the inference, as the HMM becomes a Gaussian mixture model (GMM) (see Eq. \ref{eq:transition_probability}).
Moreover, training GMM VBx on longer utterances requires significantly less memory and is several times faster than HMM VBx.

During the two-stage training, we first select the best model after the hyperparameters training according to the best validation DER. Then, we train the PLDA parameters with hyperparameters being fixed and provide results based on the best validation DER score.

\subsection{Baseline}
As a first baseline, we selected the original VBx model \cite{Landini2022VBx}, as this work directly extends it. Its hyperparameters $F_A$, $F_B$, and $P_l$ were estimated using grid search separately for each dataset to minimize DER on the development set. Label smoothing, however, was hand-tuned to $\tau = 7$ and remained constant for all datasets. 

As the HMM is simplified to a GMM in DVBx, we also include GMM VBx as a baseline. To obtain optimal results for this baseline, we re-ran the grid search for $F_A$, $F_B$.

\subsection{Evaluation}
We use oracle VAD (during training and evaluation) and 40 VB iterations with stopping criterion (i.e., stop VB if the evidence lower bound \cite{Bishop2006} stops increasing).
The models are evaluated using DER implemented in pyannote \cite{pyannote_metrics} with the standard collar=0.0\,s for AMI and DH and collar=0.25\,s for CH.

\vspace{-2mm}
\section{Results}
\label{sec:results}

\vspace{-1mm}
\subsection{Hyperparameters Fine-tuning}
\begin{figure}[tb]
\vspace{-2mm}

\centering
\centerline{\includegraphics[width=8.5cm]{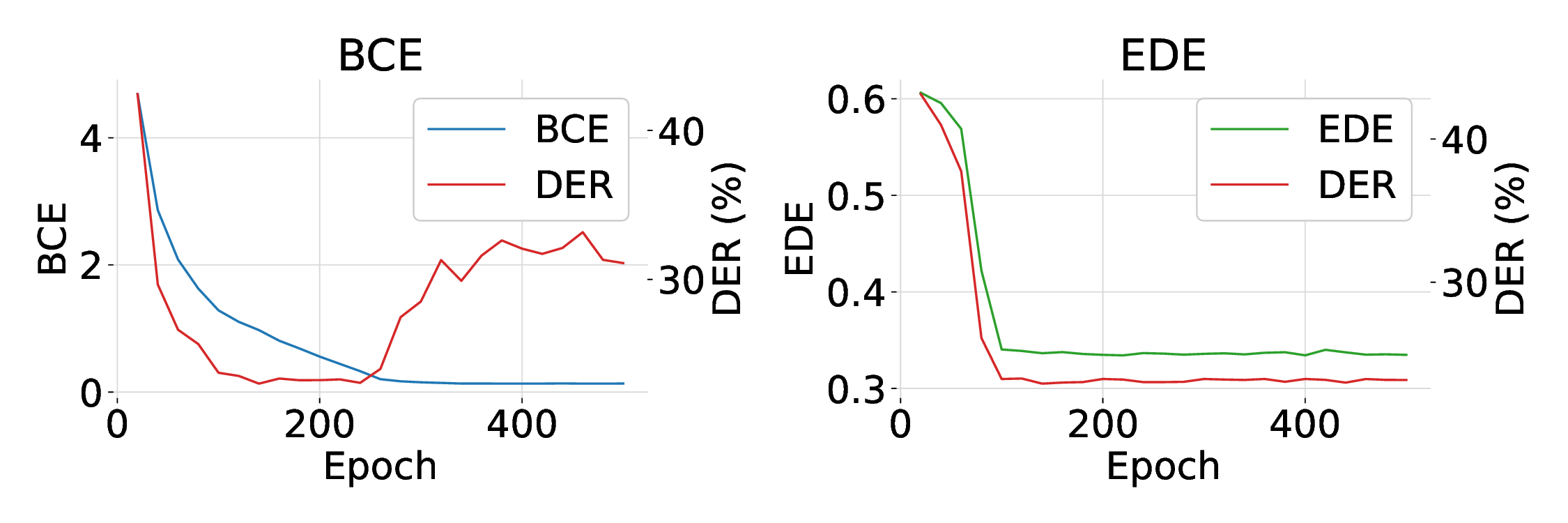}}
\vspace*{-5mm}
\caption{BCE and EDE behavior throughout 500 epochs on the CH train set.}
\label{fig:bce_der}
\vspace{-2mm}
\end{figure}

Table \ref{tab:all_results} presents the main experimental results. It first shows the performance of the baselines and their corresponding hyperparameter values. Secondly, it shows the results of hyperparameter training experiments comparing the three losses: BCE, BCE with calibration (BCE+calib.), and EDE. 

As mentioned earlier, BCE optimization and DER performance do not correlate well, which is illustrated in Fig. \ref{fig:bce_der}. BCE requires using a validation set for early stopping, as parameters would eventually diverge, significantly harming the performance. Moreover, substantially lower $F_A$, $F_B$ values are achieved by BCE because it tries to compensate for the overconfident responsibilities.
BCE+calib. solves this issue but performs worse than both BCE and EDE. The proposed EDE not only provides reasonable hyperparameter values but converges faster than BCE (see Fig. \ref{fig:bce_der}) and achieves comparable or better DER in all datasets. Hence, further comparisons are conducted on models trained with EDE.

Furthermore, Table \ref{tab:all_results} shows that DVBx with EDE outperforms the GMM VBx baseline on all datasets while achieving comparable performance with the original HMM VBx baseline, which proves that DVBx can be used to successfully tune hyperparameters. It also shows that the GMM simplification does not harm the model performance much.

\begin{table}[tb]
\caption{DERs (\%) using three losses: BCE, BCE with calibration, EDE on datasets: AMI, CH, DH. The first two rows in each dataset section display the baseline results.}
\label{tab:all_results}
\centering
\setlength{\tabcolsep}{5pt}
\begin{tabular}{@{}llcccc@{}}
\toprule
Data                & System & $\tau$ & $F_A$ & $F_B$ & DER \\
\midrule
\multirow{5}{*}{DH} & HMM VBx \cite{Landini2022VBx} & 7.00 & 0.20 & 6.00 & \textbf{18.55} \\
                    & GMM VBx  & 7.00 & 0.20 & 5.00 & 18.93 \\
\cmidrule{2-6}
                    & DVBx - BCE & 2.90 & 0.25 & 4.38 & 18.98 \\
                    & DVBx - BCE+calib. & 12.88 & 0.43 & 10.14 & 18.84 \\
                    & DVBx - EDE & 9.62 & 0.33 & 9.64 & \underline{18.76} \\
\midrule\midrule
\multirow{5}{*}{CH} & HMM VBx \cite{Landini2022VBx} & 7.00 & 0.40 & 17.00 & 13.53 \\
                    & GMM VBx & 7.00 & 0.30 & 13.00 & 13.63 \\
\cmidrule{2-6}
                    & DVBx - BCE & 0.97 & 0.08 & 1.39 & 13.53 \\
                    & DVBx - BCE+calib. & 1.93 & 0.51 & 11.16 & 14.52 \\
                    & DVBx - EDE & 12.40 & 0.26 & 9.47 & \underline{\textbf{13.48}} \\
\midrule\midrule
\multirow{5}{*}{AMI} & HMM VBx \cite{Landini2022VBx} & 7.00 & 0.40 & 64.00 & 20.84 \\ 
                    & GMM VBx & 7.00 & 0.50 & 63.00 & 21.49 \\
\cmidrule{2-6}
                    & DVBx - BCE & 12.35 & 0.12 & 8.89 & 21.06 \\
                    & DVBx - BCE+calib. & 15.10 & 0.21 & 13.90 & 21.72 \\
                    & DVBx - EDE & 3.48 & 0.25 & 25.31 & \underline{\textbf{20.91}} \\
\bottomrule
\end{tabular}
\vspace{-1mm}
\end{table}

\begin{table}[tb]
\caption{DERs (\%) of PLDA fine-tuning (FT). 
}
\label{tab:plda_results}
\centering
\setlength{\tabcolsep}{9pt} 
\begin{tabular}{@{}lccc@{}}
\toprule
System & DH & CH & AMI \\
\midrule
a)\; GMM VBx & 18.93 & 13.63 & 21.49 \\

b)\; DVBx trained $F_A,F_B,\tau$  & 18.76 & 13.48 & 20.91 \\

c)\; b) + PLDA FT & \textbf{18.66} & \textbf{13.38} &18.99 \\

\midrule\midrule
d)\; a) + PLDA FT & 18.93 & 13.63 & \textbf{18.88} \\

\bottomrule
\end{tabular}
\vspace{-2mm}
\end{table}

\vspace{-2mm}
\subsection{PLDA Fine-tuning}
PLDA fine-tuning experiments are shown in Table \ref{tab:plda_results}. It first presents the following systems: a) GMM VBx baseline, b) discriminatively trained hyperparameters, c) fine-tuned PLDA with fixed hyperparameters obtained in b).
While the fine-tuning obtains comparable performance to the baseline in DH and CH, for AMI, it yields an absolute 2.61\% DER improvement w.r.t. the GMM VBx baseline and 1.92\% DER improvement w.r.t. DVBx hyperparameters training.
We hypothesize that, in AMI, DVBx can leverage the larger development set to fix the mismatch between the pre-trained PLDA and the target domain. Lastly, for the sake of completeness, we also tested PLDA FT using the GMM VBx baseline parameters, where a similar behavior was observed.

\vspace{-2mm}
\section{Conclusion}
\label{sec:conclusion}
\vspace{-1mm}
In this work, we introduced a new principled way of tuning the VBx model hyperparameters without the need for running an extensive grid search requiring an understanding of hyperparameter sensitivity. To successfully do so, we introduced a new loss, EDE, that better correlates with the DER metric than the standard BCE loss. 
We want to highlight that while we used EDE only in the context of DVBx, its usage can be extended to EEND system training.
Furthermore, we showed that we can further improve the model performance by fine-tuning the PLDA parameters.
The presented work is the first step towards building a robust DVBx, combining the capabilities of a principled model with discriminative training. In future work, we plan to backpropagate gradients to the x-vector extractor to fine-tune together the whole diarization pipeline. Moreover, we would like to couple this method with the recently proposed well-performing MS-VBx \cite{delcroix23_interspeech}, allowing the model to handle overlaps.

\vspace{-1mm}
\section{Acknowledgements}
\label{sec:print}
\vspace{-1mm}
The work was partly supported by Czech Ministry of Interior project No. VJ01010108 "ROZKAZ", Czech National Science Foundation (GACR) project NEUREM3 No. 19-26934X, and Horizon 2020 Marie Sklodowska-Curie grant ESPERANTO, No. 101007666. Computing on IT4I supercomputer was supported by the Czech Ministry of Education, Youth and Sports through the e-INFRA CZ (ID:90254).

\vfill\pagebreak

\bibliographystyle{IEEEbib}
\bibliography{strings,refs, ref}

\begin{thebibliography}{10}

\bibitem{park2019auto}
Tae~Jin Park, Kyu~J Han, Manoj Kumar, and Shrikanth Narayanan,
\newblock ``Auto-tuning spectral clustering for speaker diarization using
  normalized maximum eigengap,''
\newblock {\em IEEE Signal Processing Letters}, vol. 27, pp. 381--385, 2019.

\bibitem{Landini2022VBx}
Federico Landini, Ján Profant, Mireia Diez, and Lukáš Burget,
\newblock ``{Bayesian HMM clustering of x-vector sequences (VBx) in speaker
  diarization: Theory, implementation and analysis on standard tasks},''
\newblock {\em Computer Speech \& Language}, vol. 71, pp. 101254, 2022.

\bibitem{horiguchi2020end}
Shota Horiguchi, Yusuke Fujita, Shinji Watanabe, Yawen Xue, and Kenji
  Nagamatsu,
\newblock ``{End-to-End Speaker Diarization for an Unknown Number of Speakers
  with Encoder-Decoder Based Attractors},''
\newblock in {\em Proc. Interspeech 2020}, 2020, pp. 269--273.

\bibitem{wang2023told}
Jiaming Wang, Zhihao Du, and Shiliang Zhang,
\newblock ``Told: a novel two-stage overlap-aware framework for speaker
  diarization,''
\newblock in {\em ICASSP 2023 - 2023 IEEE International Conference on
  Acoustics, Speech and Signal Processing (ICASSP)}, 2023.

\bibitem{bredinpyannote2.1}
Herv{\'e} Bredin,
\newblock ``Pyannote. audio 2.1 speaker diarization pipeline: Principle,
  benchmark, and recipe,''
\newblock in {\em Proceedings for Interspeech}, 2023.

\bibitem{delcroix23_interspeech}
Marc~Delcroix et.al.,
\newblock ``{Multi-Stream Extension of Variational Bayesian HMM Clustering
  (MS-VBx) for Combined End-to-End and Vector Clustering-based Diarization},''
\newblock in {\em Proc. INTERSPEECH 2023}, 2023, pp. 3477--3481.

\bibitem{raj2021multi}
Desh Raj, Zili Huang, and Sanjeev Khudanpur,
\newblock ``Multi-class spectral clustering with overlaps for speaker
  diarization,''
\newblock in {\em 2021 IEEE Spoken Language Technology Workshop (SLT)}. IEEE,
  2021, pp. 582--589.

\bibitem{yu2022m2met}
Fan Yu, Shiliang Zhang, Yihui Fu, Lei Xie, Siqi Zheng, Zhihao Du, Weilong
  Huang, Pengcheng Guo, Zhijie Yan, Bin Ma, et~al.,
\newblock ``M2met: The icassp 2022 multi-channel multi-party meeting
  transcription challenge,''
\newblock in {\em ICASSP 2022-2022 IEEE International Conference on Acoustics,
  Speech and Signal Processing (ICASSP)}. IEEE, 2022, pp. 6167--6171.

\bibitem{grauman2022ego4d}
Kristen Grauman, Andrew Westbury, Eugene Byrne, Zachary Chavis, Antonino
  Furnari, Rohit Girdhar, Jackson Hamburger, Hao Jiang, Miao Liu, Xingyu Liu,
  et~al.,
\newblock ``Ego4d: Around the world in 3,000 hours of egocentric video,''
\newblock in {\em Proceedings of the IEEE/CVF Conference on Computer Vision and
  Pattern Recognition}, 2022, pp. 18995--19012.

\bibitem{cheng2022conversational}
G.~Cheng et.al.,
\newblock ``The conversational short-phrase speaker diarization (cssd) task:
  Dataset, evaluation metric and baselines,''
\newblock in {\em 2022 13th International Symposium on Chinese Spoken Language
  Processing (ISCSLP)}. IEEE, 2022, pp. 488--492.

\bibitem{horiguchi2021end}
Shota Horiguchi, Paola Garcia, Yusuke Fujita, Shinji Watanabe, and Kenji
  Nagamatsu,
\newblock ``End-to-end speaker diarization as post-processing,''
\newblock in {\em ICASSP 2021-2021 IEEE International Conference on Acoustics,
  Speech and Signal Processing (ICASSP)}. IEEE, 2021, pp. 7188--7192.

\bibitem{singh2023supervised}
Prachi Singh, Amrit Kaul, and Sriram Ganapathy,
\newblock ``Supervised hierarchical clustering using graph neural networks for
  speaker diarization,''
\newblock in {\em ICASSP 2023-2023 IEEE International Conference on Acoustics,
  Speech and Signal Processing (ICASSP)}. IEEE, 2023, pp. 1--5.

\bibitem{snyder2018x}
David Snyder, Daniel Garcia-Romero, Gregory Sell, Daniel Povey, and Sanjeev
  Khudanpur,
\newblock ``{X-vectors: Robust DNN embeddings for speaker recognition},''
\newblock in {\em International Conference on Acoustics, Speech, and Signal
  Processing (ICASSP)}. IEEE, 2018, pp. 5329--5333.

\bibitem{brummer2010spkpartitioning}
Niko Brummer and Edward Villiers,
\newblock ``{The speaker partitioning problem},''
\newblock {\em Proc. of Odyssey 2010}, 01 2010.

\bibitem{Bishop2006}
Christopher~M. Bishop,
\newblock {\em {Pattern Recognition and Machine Learning}},
\newblock Springer-Verlag New York, Inc., Secaucus, NJ, USA, 2006.

\bibitem{yamashita2022improving}
Natsuo Yamashita, Shota Horiguchi, and Takeshi Homma,
\newblock ``Improving the naturalness of simulated conversations for end-to-end
  neural diarization,''
\newblock in {\em Odyssey 2022}, 2022.

\bibitem{landini2022InterDataEEND}
Federico Landini, Alicia Lozano-Diez, Mireia Diez, and Lukáš Burget,
\newblock ``{From Simulated Mixtures to Simulated Conversations as Training
  Data for End-to-End Neural Diarization},''
\newblock in {\em Proc. Interspeech}, 2022, pp. 5095--5099.

\bibitem{burget2011discriminatively}
Luk{\'a}{\v{s}} Burget, Old{\v{r}}ich Plchot, Sandro Cumani, Ond{\v{r}}ej
  Glembek, Pavel Mat{\v{e}}jka, and Niko Br{\"u}mmer,
\newblock ``Discriminatively trained probabilistic linear discriminant analysis
  for speaker verification,''
\newblock in {\em International Conference on Acoustics, Speech, and Signal
  Processing (ICASSP)}. IEEE, 2011, pp. 4832--4835.

\bibitem{McCree2014MulticlassDT}
Alan McCree,
\newblock ``Multiclass discriminative training of i-vector language
  recognition,''
\newblock {\em The Speaker and Language Recognition Workshop (Odyssey 2014)},
  2014.

\bibitem{yu2017pit}
Dong Yu, Morten Kolbæk, Zheng-Hua Tan, and Jesper Jensen,
\newblock ``Permutation invariant training of deep models for
  speaker-independent multi-talker speech separation,''
\newblock in {\em 2017 IEEE International Conference on Acoustics, Speech and
  Signal Processing (ICASSP)}, 2017, pp. 241--245.

\bibitem{fujita2019end}
Yusuke Fujita, Naoyuki Kanda, Shota Horiguchi, Kenji Nagamatsu, and Shinji
  Watanabe,
\newblock ``{End-to-End Neural Speaker Diarization with Permutation-Free
  Objectives},''
\newblock in {\em Proc. Interspeech 2019}, 2019, pp. 4300--4304.

\bibitem{DIHARD19}
N.~Ryant et. al.,
\newblock ``{The Second DIHARD Diarization Challenge: Dataset, task, and
  baselines.},''
\newblock in {\em Proceedings of Interspeech}, 2019.

\bibitem{Callhome}
A.~F. Martin and M.~A. Przybocki,
\newblock ``{Speaker recognition in a multi-speaker environment},''
\newblock in {\em 7th European Conference on Speech Communication and
  Technology, Eurospeech}, September 2001, vol. 7, num. 2, pp. 787--790.

\bibitem{carletta2005ami}
Jean Carletta, Simone Ashby, Sebastien Bourban, Mike Flynn, Mael Guillemot,
  Thomas Hain, Jaroslav Kadlec, Vasilis Karaiskos, Wessel Kraaij, Melissa
  Kronenthal, et~al.,
\newblock ``{The AMI meeting corpus: A pre-announcement},''
\newblock in {\em International workshop on machine learning for multimodal
  interaction}. Springer, 2006, pp. 28--39.

\bibitem{he2016deep}
Kaiming He, Xiangyu Zhang, Shaoqing Ren, and Jian Sun,
\newblock ``{Deep residual learning for image recognition},''
\newblock in {\em Proceedings of the IEEE conference on computer vision and
  pattern recognition}, 2016, pp. 770--778.

\bibitem{zeinali2019but}
Hossein Zeinali, Shuai Wang, Anna Silnova, Pavel Mat\v{e}jka, and Old\v{r}ich
  Plchot,
\newblock ``But system description to voxceleb speaker recognition challenge
  2019,''
\newblock in {\em Proceedings of The VoxCeleb Challange Workshop 2019}, 2019,
  pp. 1--4.

\bibitem{adam_optim}
Diederik~P. Kingma and Jimmy Ba,
\newblock ``Adam: {A} method for stochastic optimization,''
\newblock in {\em 3rd International Conference on Learning Representations,
  {ICLR}}, 2015.

\bibitem{pyannote_metrics}
Herv\'e Bredin,
\newblock ``{pyannote.metrics: a toolkit for reproducible evaluation,
  diagnostic, and error analysis of speaker diarization systems},''
\newblock in {\em {Proceedings of Interspeech 2017}}, 2017.

\end{thebibliography}

\end{document}